\documentclass[aps,pra,superscriptaddress,notitlepage,twocolumn, 10pt,longbibliography]{revtex4-1}
\usepackage{hyperref}
\usepackage{amsmath}
\usepackage{graphicx}
\usepackage{color}
\usepackage{braket}
\usepackage{epstopdf} 
\usepackage{subcaption}

\usepackage{bm}

\newcommand{\addcqt}{Centre for Quantum Technologies, National University of Singapore, 3 Science Drive 2, Singapore 117543}

\newcommand{\addsutd}{Science, Mathematics and Technology Cluster, Singapore University
of Technology and Design, Singapore}
\begin{document}

\title{Cusp solitons mediated by a topological nonlinearity}


\author{Harvey Cao}
\affiliation{\addcqt}

\author{Daniel Leykam}
\affiliation{\addsutd}

\date{\today}

\begin{abstract}
    Nonlinearity in the Schrödinger equation gives rise to rich phenomena such as soliton formation, modulational instability, and self-organization in diverse physical systems. Motivated by recent advances in engineering nonlinear gauge fields in Bose-Einstein condensates, we introduce a nonlinear Schrödinger model whose dynamics are dependent on the curvature of the wavefunction intensity and show that this has a direct link to a topological quantity from persistent homology. Our model energetically penalizes or favours the formation of local extrema and we demonstrate through numerical simulations that this topological nonlinearity leads to the emergence of robust, cusp-like soliton structures and supports flat-top beams which do not suffer from conventional modulational instability. These findings suggest that topological nonlinearities could serve as a versatile tool for controlling nonlinear waves in optics and Bose-Einstein condensates.

\end{abstract}

\maketitle

{\bf Nonlinear waves, such as those observed in optical fibres and quantum gases, have long interested physicists for their ability to form stable patterns in systems governed by the nonlinear Schrödinger equation (NLSE). Recent advances in engineering quantum materials and photonic devices have enabled the creation of new kinds of nonlinear responses that are linked to the shape of wave patterns. In this article, we introduce and analyze a novel NLSE model whose ``topological nonlinearity" penalizes or favours the formation of peaks and dips in wave intensity based on its curvature. This leads to the emergence of robust, cusp-like soliton solutions and dynamical stabilization of flat-top beams, which usually disappear in the conventional NLSE setting due to nonlinear instabilities. Our results reveal a new pathway for controlling nonlinear systems in optical and quantum systems, as well as bridging the gap between concepts from applied topology and nonlinear physics.
}

\section{Introduction}

The nonlinear Schrödinger equation (NLSE) appears throughout physics to model systems where wavelike self-interactions occur. When this nonlinearity is balanced by dispersion, it gives rise to interesting phenomena such as modulational instability and the formation of stable, non-dispersive topological structures known as solitons. These self-stabilizing solutions are widely studied for both their mathematical properties as well as their considerable potential applications in information and communications technology \cite{PhysRevE.63.016610, PhysRevE.64.016612, Krolikowski2004, TURITSYN20021741, porsezian2003optical}. 

Such behaviour has been found to be universal and solitons have been identified in a wide range of physical systems: within the field of ultracold atomic physics, the NLSE describes Bose-Einstein condensates (BECs) in the limit of a dilute gas of weakly-interacting bosons; in optical systems, the NLSE describes light propagation through nonlinear media, resulting in behaviour such as self-focusing beams; solitons also feature in the study of plasmas under certain conditions \cite{PhysRevLett.94.040403, Krolikowski2004,porkolab1976upper, 10.1007/3-540-09246-3_9}.

Since the initial investigations of the NLSE, a wide range of studies have explored its diverse generalizations, revealing a variety of nonlinear phenomena and an understanding of the criteria for soliton stability \cite{PhysRevE.89.012907, PhysRevA.72.033611}. In particular, extensions to incorporate higher-order and nonlocal nonlinearities have demonstrated that solitons can be stabilized against dynamical collapse under appropriate conditions \cite{PhysRevE.66.046619,PhysRevA.77.043821,PhysRevE.73.055601,RABIE2024107589}. Crucially, these nonlocal nonlinear responses are intrinsically linked to the local geometry and curvature of the wavefunction intensity, which enables new mechanisms for controlling nonlinear excitations that are not achievable with purely local nonlinearities. 

Further theoretical work on topologically-motivated nonlinearities has demonstrated the formation of topological solitons and opened new perspectives for quantum simulations \cite{edmonds_2013, PhysRevA.92.013604, BISWAS20092845}. In particular, it has been shown that density-dependent gauge fields in the discrete NLSE can promote the formation of stable domain walls between low and high density regions - topological defects - whose dynamical response to the gauge field is observed to be drastically different from the bare atoms \cite{PhysRevLett.132.023401, PhysRevE.103.032206, YAO2022}. Parallel experimental advances have shown how these exotic forms of nonlinearities can be realized using schemes such as Floquet modulation based periodic modulation of the lattice potential and nonlinearity strength
\cite{PhysRevLett.121.030402, Martinez_2016,PhysRevLett.121.030402,Gorg_2019}.

Motivated by these recent works on nonlinear gauge fields, in this work we introduce and analyze a nonlocal nonlinear Schr\"odinger model inspired by persistent homology, which is an unsupervised machine learning method for identifying significant topological features within a dataset. The considered nonlocal nonlinearity is proportional to the sign of the gradient of the intensity. After a suitable transformation, this nonlocal nonlinearity corresponds to an energy penalty or reduction for each local extremum of the intensity. Similar nonlinearities may arise in optics via light-induced gradient forces acting on small particles~\cite{PhysRevA.75.055801,PhysRevLett.105.163906,Gautam2019}. We investigate the implications of this peculiar purely nonlocal nonlinearity, which leads to the formation of cusp solitons with different topological features depending on the sign and strength of the nonlinearity.

The outline of this paper is as follows: Section II provides an introduction to the our NLSE model. Section III draws a connection to sublevel set filtration persistent homology. In Section IV, we outline the propagation results from three types of initial beam profile and discuss the potential implications. Section V makes concluding remarks and suggests future research directions. An Appendix outlines the techniques used for numerical simulation and analysis of the model and its ground states. 

\section{Model}

We consider the one-dimensional nonlocal NLSE. Nonlocal NLSEs arise in various contexts, particularly nonlinear optics, and can be written generically in dimensionless units as 
\begin{equation}
i \partial_t \psi = -\partial_x^2 \psi + N[|\psi|^2] \psi, \label{eq:nlse}
\end{equation}
where the wave field $\psi(t,x)$ simultaneously undergoes diffraction (described by the $\partial_x^2$ term) and interaction with an nonlinear potential $N[|\psi|^2]$ dependent on the field intensity profile $|\psi(x)|^2$. One commonly-employed model for nonlocal nonlinearity is
\begin{equation}
N[I] = g\int dx^{\prime} R( x^{\prime} - x) |\psi(x^{\prime})|^2, \label{eq:nonlocal}
\end{equation}
where the nonlocal response function $R$ describes how the potential shift depends on the intensity profile and the parameter $g$ determines the strength and sign of the nonlinearity. For example, in the case of thermal nonlinearities, heat diffusion will tend to smear out the induced potential to a finite size determined by the diffusion constant of the medium.

When the response is short-ranged, it can be approximated by a $\delta$ function, $R(x) = g \delta(x)$, which yields the standard local NLSE,
\begin{equation}
i \partial_t \psi = -\partial_x^2 \psi + g |\psi|^2 \psi.
\end{equation}
The simplest nonlocal correction to this equation is to assume $R$ is non-singular but strongly localized, such that the width of the nonlocal response is much less than the width of the beam $\psi$. In this case one can expand the nonlocal response function as
\begin{equation}
N[|\psi|^2] \approx g ( 1 + \gamma \partial_x^2) |\psi|^2, \label{eq:weak}
\end{equation}
where the parameter $\gamma = \frac{1}{2}\int dx \, x^2R(x)$ characterizes the strength of the nonlocality \cite{Krolikowski2004} and is independent of the fine details of the nonlocal response function. In this weakly nonlocal limit there is an additional shift to the effective potential, proportional to the curvature of intensity, which affects the width and power thresholds of bright and dark solitons~\cite{PhysRevE.63.016610,PhysRevE.64.016612}.

To obtain a topological nonlinearity, we shall neglect the local term in Eq.~\eqref{eq:weak} and replace the curvature term with a step-like response, namely
\begin{equation} 
N[|\psi|^2] = \alpha \partial_x \mathrm{sgn}(\partial_x |\psi|^2) \approx \alpha \partial_x \left[ \tanh \left(\frac{\partial_x |\psi|^2}{w} \right) \right], \label{eq:topo-nl}
\end{equation}
where $w$ is a regularization parameter. The limit $w \rightarrow 0$ yields the sign function response, while $w \rightarrow \infty$ reproduces the weakly nonlocal limit of Eq.~\eqref{eq:weak} with $\gamma = \alpha / w$. For a uniform intensity profile, this nonlinearity vanishes, whereas non-uniform fields experience energy shifts at local extrema dependent on the curvature of the intensity.

Physically, nonlocal response functions similar Eq.~\eqref{eq:topo-nl} may arise for nonlinearities mediated by small particles suspended in a fluid and accelerated by optical gradient forces, leading to peaks in the wavefunction intensity at local extrema of the intensity where $\partial_x |\psi|^2 = 0$~\cite{PhysRevA.75.055801,PhysRevLett.105.163906,Gautam2019}. The regularization parameter $w$ then accounts for a finite spread of the particles about the extrema due to repulsion or other effects. In the following, we will consider the limit $w \rightarrow 0$ for our analytical results and use a small finite $w$ in numerical simulations to avoid numerical instabilities.

\section{Topological perspective}

The conserved energy associated with the topological nonlinearity Eq.~\eqref{eq:topo-nl} can be found as
\begin{align}
    E[\psi,\psi^{*}] &= \int_{-\infty}^{\infty} |\partial_{x}\psi|^{2} dx - \int_{-\infty}^{\infty} \alpha \,\text{sgn}(\partial_{x}|\psi|^{2})(\partial_{x}|\psi|^{2})dx ,  \nonumber \\
    &\equiv E_{L} - E_{NL}, 
    \label{eq:energy}
\end{align}
which we have written as an integral over the entire field profile. Here, we show that it is possible to view the nonlinearity from a topological perspective, in which the nonlinear energy shift $E_{NL}$ depends only on the local maxima and minima of intensity and not on the precise details of the intensity distribution.

To do so, we assume a localized (normalizable) field profile with $\psi \rightarrow 0$  as $|x| \rightarrow \infty$ and evaluate $E_{NL}$ by splitting the integral into intervals where $\partial_x |\psi|^2$ is strictly increasing or decreasing, corresponding to $\mathrm{sgn}(\partial_x |\psi|^2) = \pm 1$. Evaluating the $E_{NL}$ integral from the first minimum ($b_1 =0$ at $x=-\infty$) to the first local maximum, we obtain
\begin{equation}
\int^{x_{1}}_{-\infty}\, \partial_{x}|\psi|^{2} dx  = |\psi(x_1)|^2 \equiv d_1 - b_1,
\end{equation} 
since $\mathrm{sgn}(\partial_x |\psi|^2) = 1$ in this interval. Continuing the integral up to the next local minimum,
\begin{align}
    \int_{x_1}^{x_2} \mathrm{sgn} (\partial_x |\psi|^2) \partial_x |\psi|^2 dx &= - \int_{x_1}^{x_2} \partial_x |\psi|^2 dx, \nonumber  \\
    &= d_1 - |\psi(x_2)|^2, \nonumber \\ 
    &\equiv d_1 - b_2,
\end{align}
followed by the next local maximum,
\begin{align}
    \int_{x_2}^{x_3} \mathrm{sgn} (\partial_x |\psi|^2) \partial_x |\psi|^2 dx &= \int_{x_2}^{x_3} \partial_x |\psi|^2 dx, \nonumber  \\
    &= |\psi(x_3)|^2 - b_2, \nonumber  \\
    &\equiv d_2 - b_2,
\end{align}
and so on. Thus, we can write the nonlinear contribution to the conserved energy as a sum of the intensities at the local maxima ($d_i$), minus the intensities at the local minima ($b_i$),
\begin{equation}
    E_{NL} = 2\sum_i (d_i - b_i).
\end{equation}
Remarkably, the same quantity emerges as a topological summary statistic when applying persistent homology to quantify the shape of the intensity profile $|\psi(x)|^2$. Specifically, persistent homology considers the sublevel set of points X where a real-valued function $f(x)$ is below a certain threshold $\lambda$
\begin{equation}
    X = f^{-1}(-\infty, \lambda]
\end{equation}
By varying $\lambda$, a sequence of nested topological spaces known as a filtration is generated. Observing the change in topological features throughout this sequence provides a quantitative topological summary of the whole data set. This provides an indication as to which topological features (such as connected components, cycles, and higher-dimensional holes) are more or less significant based on how long they persist through the filtration~\cite{10.1145/2535927}. We note that the birth and death of connected components correspond to local maxima and local minima of the function, respectively. 

The homology of a sublevel set filtration only changes when critical points of the function are swept across, such as minima, maxima, and saddle points, as shown in Fig. 1. For quantum physics, persistent homology has been particularly useful for studying properties of energy landscapes and quantum states, offering insights into phase transitions, entanglement, and the behaviour of systems under external perturbations \cite{murugan2019introductiontopologicaldataanalysis, Leykam31122023, PhysRevB.107.115174, PhysRevE.107.044204, Hamilton2024, PhysRevE.93.052138}.

\begin{figure}
  \centering {\includegraphics[width=\columnwidth]{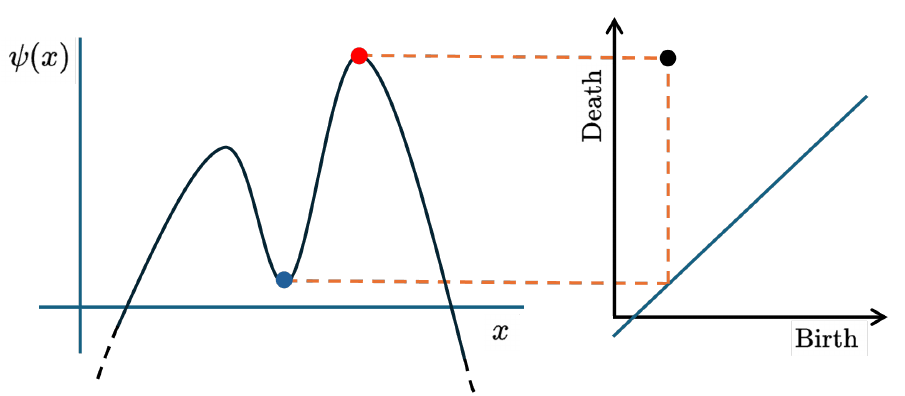}}%
  \caption{Illustration of sublevel set filtration persistent homology from a 1D profile. One example persistent feature (black dot) is shown, which is born at the local minimum (blue dot) and dies at the subsequent local maximum (red dot).}
  \label{fig:fig1}
\end{figure}

In persistent homology, it is common to summarise the persistence of topological features (in this case, the local maxima and minima) by computing a norm of the feature lifetimes, e.g. \cite{CohenSteiner2010, PhysRevC.106.064912,PhysRevB.104.104426, PhysRevE.93.052138, pmlr-v139-carriere21a, murugan2019introductiontopologicaldataanalysis}
\begin{equation}
    P_{p} = (\sum_{i}(d_{i}-b_{i})^{p})^{\frac{1}{p}}, 
\end{equation}
where the summation is made over all features and the choice of $p$ affects the sensitivity to the persistence of features. The case $p=1$, corresponding to the sum of feature lifetimes $l_{i} = d_{i} - b_{i}$, is precisely the nonlinear energy term appearing in Eq.~\eqref{eq:energy}. Thus, our model introduces a novel line of research which incorporates for the first time quantities deriving from persistent homology directly into quantum dynamics.

\section{Dynamics}

In order to understand the propagation characteristics of our model, we use the split-step method to simulate beam propagation from a few different initial profiles. The evolution of the intensity profiles in both spatial and Fourier space will reveal how dispersion and nonlinearity interact to shape the beam dynamics, including features such as self-focusing or defocusing, soliton formation, and the development of spatial or spectral instabilities, depending on the sign and strength of the nonlinearity. Here, we consider three types of initial profile - a Gaussian beam, a dark soliton-type beam, and a flat-top beam. 

\subsection{Gaussian beam}

\begin{figure*}
\centering
\includegraphics[width=2\columnwidth]{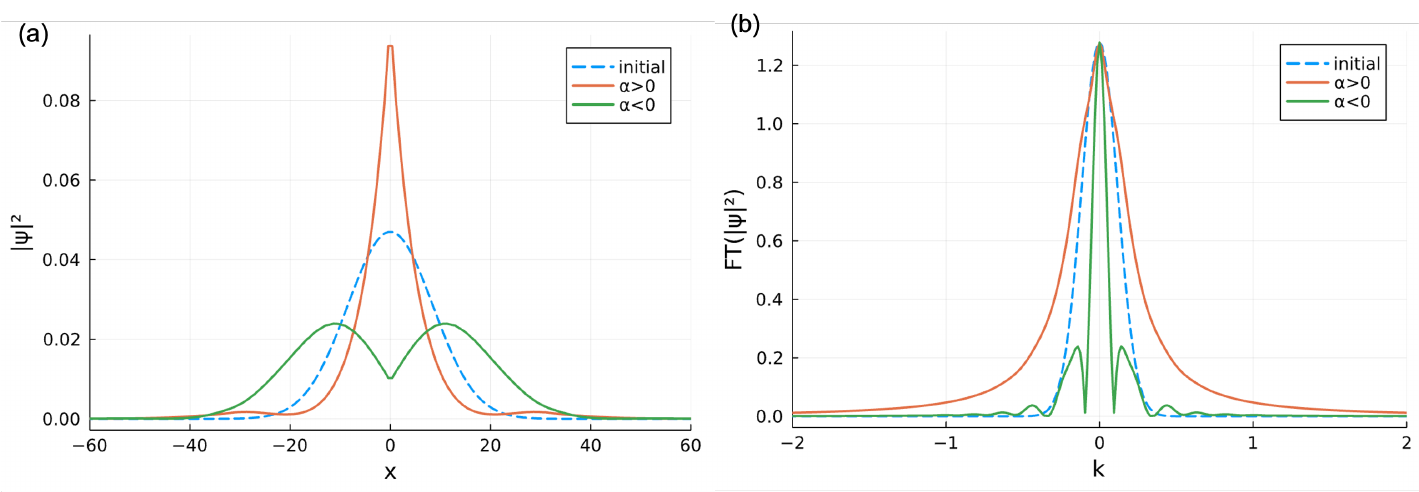}
\caption{Beam propagation ($T=500$) of an initial Gaussian profile (blue dotted line) under the topological nonlinearity model in (a) real and (b) Fourier space. For $\alpha>0$ (red line), the beam undergoes a sharpening effect, whereas for $\alpha <0$ (green line) it forms a strong dip, splitting the beam into two.}
\label{fig:fig2}
\end{figure*}

\subsubsection{Self-focusing regime}

The Gaussian beam serves as a canonical initial condition for probing the interplay between dispersion and nonlinearity in both the conventional NLSE and our persistent homology-inspired model. In the standard cubic NLSE, the evolution of a Gaussian profile is well understood: for focusing nonlinearity, self-interactions counteract diffraction, leading to self-focusing and, above a critical power threshold, the formation of bright solitons - spatially localized, non-dispersive wave packets \cite{MA2013718, KARJANTO2022}.

In our model, for $\alpha > 0$, the nonlinearity acts as an effective attractive potential at points of positive curvature in the intensity profile. Thus an initial Gaussian beam rapidly self-focuses, forming a highly localized structure with a pronounced cusp at its center and slow attenuation towards infinity (see Fig. 2a). The corresponding Fourier spectrum in Fig. 2b is Lorentzian in form, which would be consistent with the $\exp(-w|x|)$ real space profile, since the Fourier transform of an exponential decay produces a Lorentzian lineshape. This spectral signature contrasts with the Gaussian Fourier spectrum of the initial condition, highlighting the nonlinearity-induced reshaping. These kinds of Fourier spectra are relevant for optical lattice realizations with time-of-flight imaging, which allows measuring the intensity profile in $k$-space directly \cite{PhysRevA.77.043626}.

\subsubsection{Defocusing regime}

In contrast, the defocusing regime of the conventional NLSE enhances the diffraction of Gaussian beams and can result in the formation of dark solitons. These are usually finite intensity dips on a uniform background, accompanied with a phase shift across the dip.

For our model, we notice that for $\alpha < 0$, where the potential is now repulsive at local maxima. Thus, the initial intensity maximum at the beam centre is transformed to a local minimum, with two peaks appearing on either side. Notably, the topology of the sublevel set filtration is altered under this regime since the number of connected components (0-D topological features) is increased; the initial single peak beam is transformed to a double-peaked beam. 

\subsection{Dark soliton state}

Dark solitons are fundamental nonlinear excitations of the defocusing NLSE, characterized by localized intensity dips embedded in a continuous nonzero background field. Given the structural similarity between the self-focusing soliton solutions of the conventional cubic NLSE and those of our model, we investigate whether Gaussian-shaped dark solitons exhibit analogous behavior under the influence of the topological nonlinearity.

The numerical simulations reveal that the evolution of dark solitons in our model qualitatively parallels that of Gaussian beams: for one sign of the nonlinear parameter, corresponding to effective self-focusing of Gaussian beams, the local curvature of the intensity profile is enhanced, reinforcing the dip structure. Thus, the same nonlinearity sign supports bright cusp solitons and dark soliton-like states on a uniform background. Conversely, for the opposite sign, the local minimum in intensity inverts into a local maximum. 

\begin{figure}
  \centering {\includegraphics[width=\columnwidth]{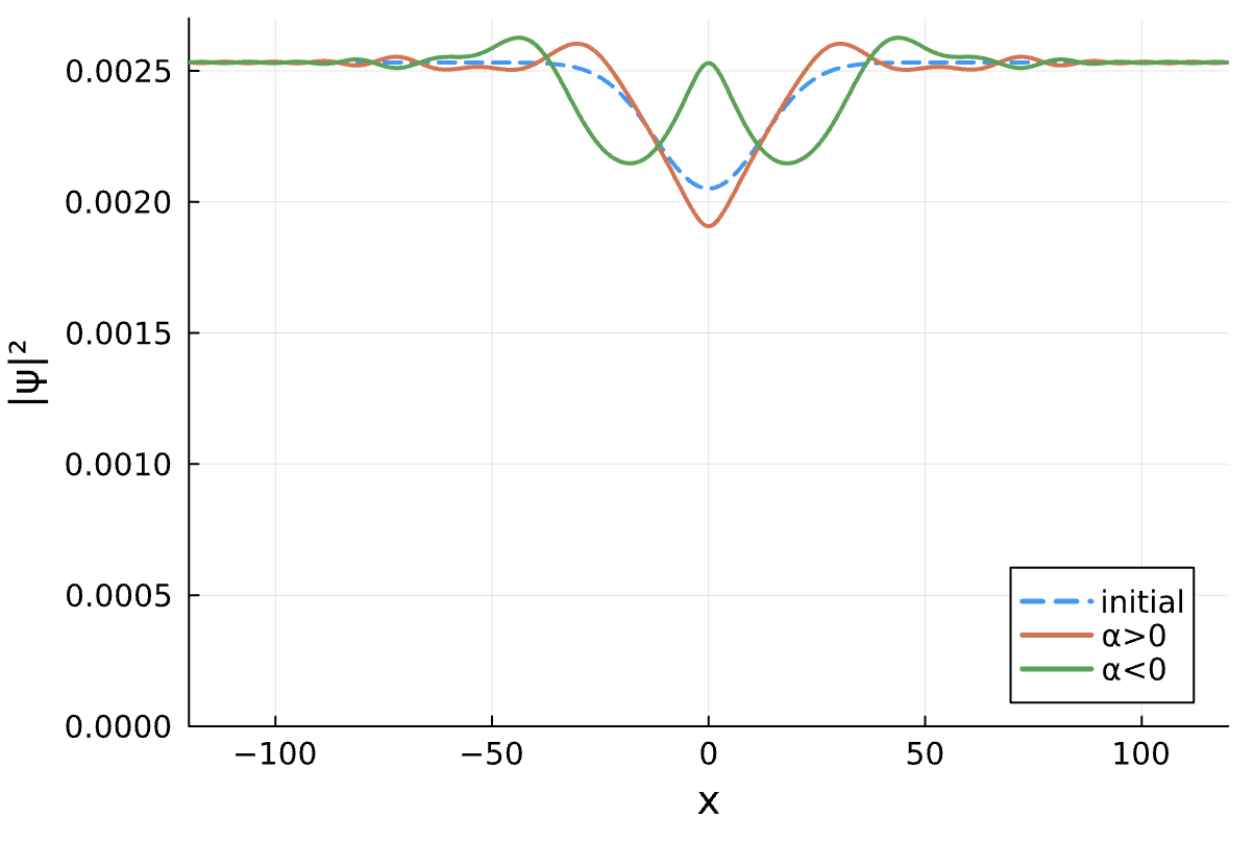}}%
  \caption{Beam propagation ($T=500$) of an initial Gaussian-shaped dark soliton (blue dotted line) under the topological nonlinearity model in real space. The behaviour is qualitatively similar as for the Gaussian initial profile in Fig. 2, whereby the peak is either sharpened (red line) or inverted (green line) depending on the sign of $\alpha$.}
  \label{fig:fig3}
\end{figure}

\subsection{Flat-top beam}

\begin{figure*}
  \centering {\includegraphics[width=2\columnwidth]{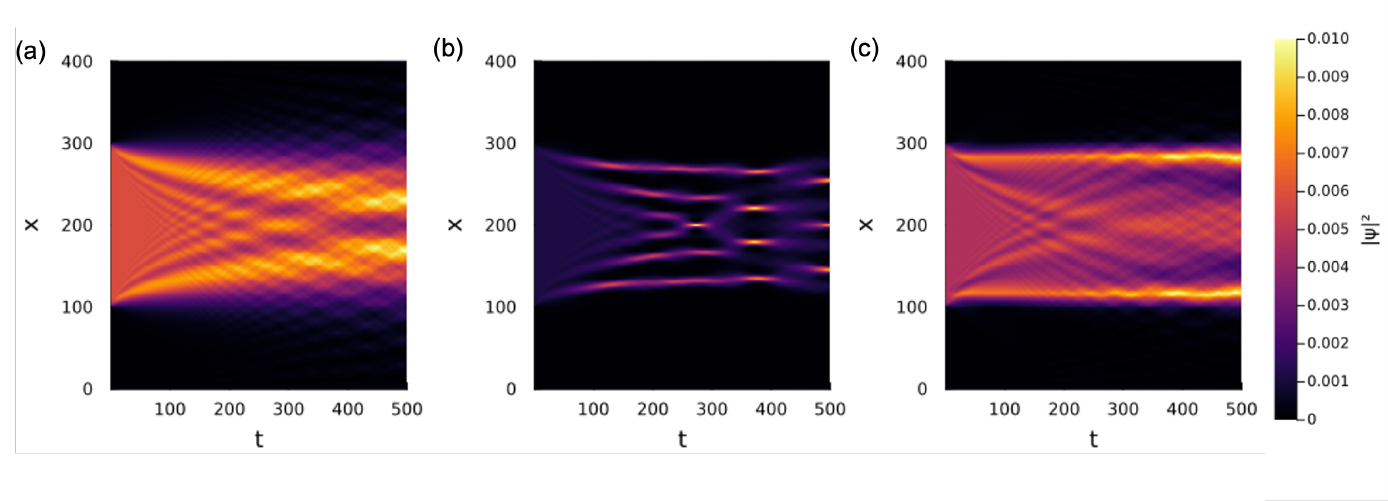}}%
  \caption{Beam propagation of an initial flat-top profile under the linear (left), cubic nonlinear (middle) and topological nonlinear (right) Schrödinger models in real space. Notably, the propagation under the topological nonlinear model differs from the linear and cubic nonlinear cases as it maintains the sharp edges of the flat-top profile.}
  \label{fig:fig4}
\end{figure*}

In the conventional NLSE, the stability of a uniform solution is determined by the sign of the nonlinearity. In the defocusing regime, the uniform state is stable; in the focusing regime, however, the uniform solution is subject to modulational instability (MI), whereby small perturbations are exponentially amplified, eventually leading to the formation of localized structures such as solitons or breathers \cite{PhysRevE.57.3510}.

In our model, the uniform state remains a trivial solution, since the absence of local extrema in $|\psi|^2$ means the topological energy penalty vanishes. To probe the stability of this solution further, we consider a wide flat-top beam with sharp transitions at the edges, a configuration often used to investigate boundary-induced instabilities and soliton formation. It is also experimentally relevant as flat-top beams serve as finite-domain approximations for infinite uniform states \cite{CHEN2025112776, ZHENG2021105016}. Under the conventional NLSE, such a beam typically evolves as follows: for defocusing nonlinearity, the flat top gradually disperses and approaches a Gaussian-like profile as the plateau narrows and the edges smooth out; for focusing nonlinearity, MI leads to the rapid formation of localized spikes and the eventual breakup of the plateau (see Fig. 4, middle). In the linear case where $g=0$, the beam will also lose its flat-top shape as it propagates due to strong diffraction at the edges, evolving towards a smoother, bell-shaped distribution (see Fig. 4, left).

In contrast, in our topological model, the evolution is qualitatively distinct. The sharp edges of the flat-top beam are preserved, and we observe the emergence of stable, localized features (peaks or dips) just inside the beam boundaries, depending on the sign of $\alpha$ ($\alpha>0$ shown in Fig. 4 right). These edge features remain stationary and robust throughout the simulation, indicating a suppression of MI and a stabilization of the plateau against both long- and short-wavelength perturbations. This behavior can be understood as a consequence of the topological energy term, which penalizes the creation of new local extrema and thus energetically disfavors the amplification of small fluctuations that would otherwise seed MI.

The stabilization of flat-top beams by a topological nonlinearity has several notable implications. First, it suggests a new mechanism for controlling the stability of nonlinear systems and suppressing MI using geometric and topological constraints. Secondly, the persistent edge features may find application in edge detection for optical systems or in machine learning, where the natural enhancement and stabilization of edge features could serve as a physics-inspired preprocessing step for identifying boundaries in structured data. 

\subsection{Ground states}

\begin{figure}
  \centering {\includegraphics[width=\columnwidth]{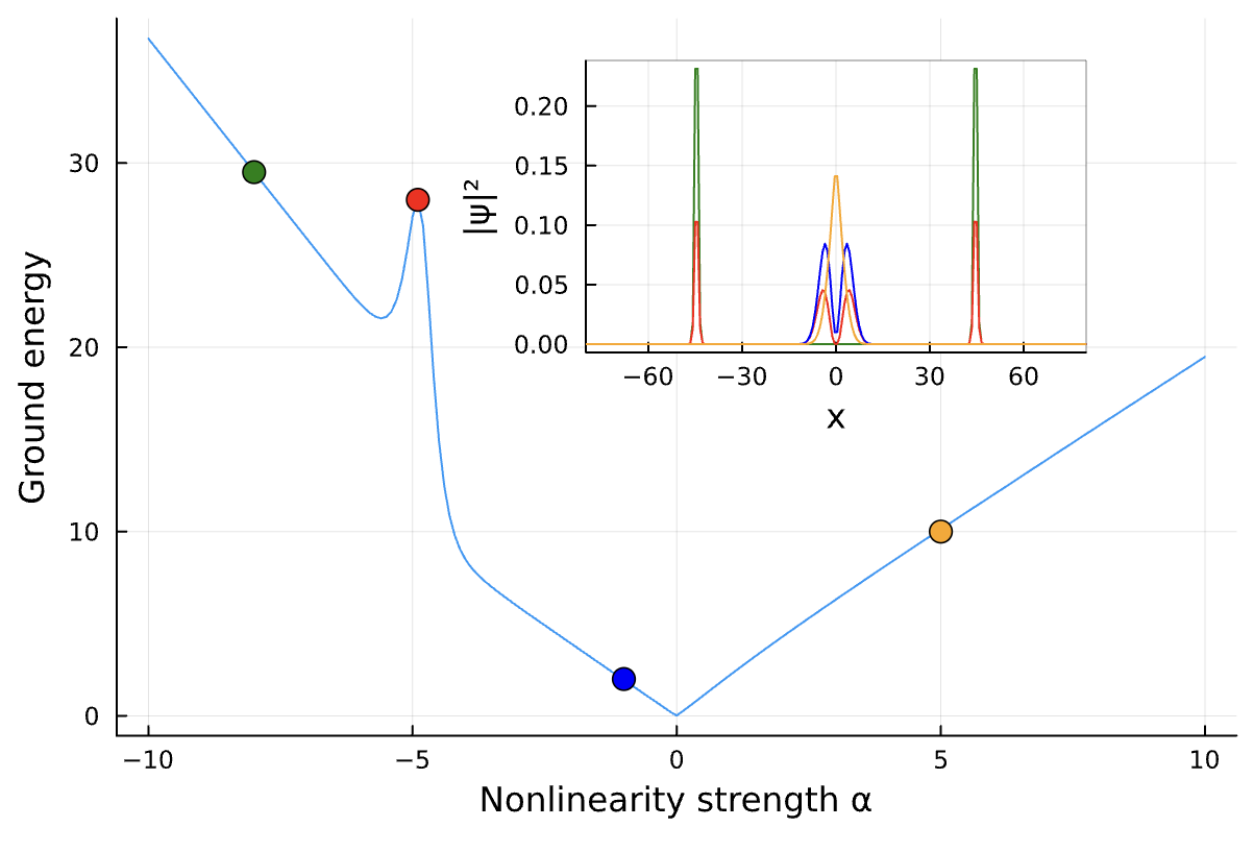}}%
  \caption{Relationship between ground state energy and the nonlinearity strength $\alpha$ for the topological NLSE (blue curve). The coloured points identify distinct ground state phases, the intensity profiles for which are shown in the inset figure with corresponding colours.}
  \label{fig:fig5}
\end{figure}

To corroborate the beam propagation simulation results, we numerically obtain the ground states of our model as a function of the nonlinearity strength $\alpha$. Starting from a normalized Gaussian trial wavefunction, we propagate the system in imaginary time $\tau$ using a symmetrized split-step ITE method. The spatial domain is discretized over $N=1024$ grid points with spacing $\Delta x = 0.01$ and system size $L=400$, ensuring negligible boundary effects. Similar to the bright solitons in the conventional focusing NLSE, we observe sharply localized cusp-like solitons for $\alpha>0$, where the topological term energetically favours the formation of local extrema. Conversely, for $\alpha < 0$, the energy landscape favors the proliferation of intensity minima, leading to bifurcated or multi-peak structures, reminiscent of the delocalized or dark soliton regimes in the defocusing NLSE.

Analytically, we can obtain some qualitative insights to the behavior of our model through seeking a self-consistent ansatz. Assuming an exponential profile $\psi(x) = A \, \text{exp}(-\kappa|x|)$, where $1/\kappa$ is the localization length, we try to find a self-consistent solution to the nonlinear wave equation. Consider a potential of the form $V(x) = \alpha \delta(\partial_{x}|\psi|^2)\partial_{x}^2$, which generates a delta function at $x=0$ for the exponential profile. In the regions where $|x|>0$, the potential is zero, and we find from substituting into the wave equation and choosing $A$ such that the wavefunction is bounded and continuous, $E = -\frac{\kappa^2}{2} < 0$. By integrating the Schrodinger equation through the delta function discontinuity at $x=0$, we find a relationship between the amplitude and localization strength, suggesting a self-consistent solution of the form $\psi(x) = \sqrt{\alpha} e^{-\alpha|x|}$. These cusp solitons have an energy $E<0$, indicating that they are favoured over the (linear) $k=0$ plane wave ground state. Similar cusp solitons also exist as exact solutions of the nonlinear Schrodinger equation with a purely nonlocal nonlinearity~\cite{Krolikowski2004,porkolab1976upper}.

\section{Conclusion}

In this study, we have introduced a novel nonlinear Schrödinger-type model whose nonlinearity is explicitly dependent on the local curvature of the field intensity. Whilst the form of soliton solutions is similar to those observed in the conventional NLSE, our model captures nontrivial geometric effects absent in amplitude-based formulations. Our analysis shows that the inclusion of a topological energy penalty, derived from the persistent homology of the intensity profile, fundamentally alters both the stability and the dynamical evolution of localized wave packets. Specifically, we find that the sign-dependent nonlinearity leads to the emergence of stable cusp-like soliton solutions, modifies the modulational stability of flat-top beams, and can promote the formation of multi-peak structures through topological bifurcations. These effects are robust to numerical regularization and persist across a range of initial conditions and feasible parameter regimes.

Specifically, in the dimensionless NLSE Eq.~\eqref{eq:nlse}, $x$ is measured in units of some characteristic length scale $a$, time is measured in units of $\tau = 2 m_{\mathrm{eff}} a^2 / \hbar$, where $m_{\mathrm{eff}}$ is the particle effective mass. The Bose-Einstein condensate with a synthetic gauge field experiment of Ref.~\cite{YAO2022} used $^133$Cs atoms with $m_{\mathrm{eff}} = 1.8 \times 10^{-25}$ kg; taking $a = 1 \mu$m yields $\tau \approx 0.5$ms. Therefore, the results in Figs.~\ref{fig:fig2} -- \ref{fig:fig4} correspond to an evolution time of about 300 ms, within the measured 1/e lifetime of $700$ ms under periodic driving. On the other hand, experiments on optical force-induced nonlinearities and self-focusing~\cite{Gautam2019} have used a green laser with a wavelength $\lambda = 532$ nm, with $m_{\mathrm{eff}} = 2\pi \hbar/\lambda$, leading to $\tau \approx 0.02$ mm, meaning our simulations correspond to feasible propagation lengths of about 1 cm. 

By bridging concepts from computational topology and nonlinear physics, our work opens new conceptual and practical avenues for the controlled manipulation of solitons, the stabilization of nonlinear modes, and the realization of topologically protected states in photonic and optical systems. Future research may extend these ideas to higher dimensions, explore the interplay with disorder and external potentials, and seek experimental realizations in optical lattices or Bose-Einstein condensates. Our findings suggest that topological nonlinearities could serve as a versatile tool for both fundamental studies and technological applications in nonlinear quantum systems.

\section*{Acknowledgments}

This research is supported by the National Research Foundation, Singapore and A*STAR under its CQT Bridging Grant and Quantum Engineering Programme NRF2021-QEP2-02-P02, A*STAR (\#21709) and by EU HORIZON- Project 101080085 — QCFD. D.L. acknowledges support from the Ministry of Education of Singapore under its  SUTD Kickstarter Initiative (Grant No. SKI 20210501).

\section*{Author Declarations}

\subsection{Conflict of Interest}
The authors have no conflicts to disclose.
\subsection{Data availability}
The code used to perform the numerical calculations
within this paper is available from the corresponding author upon reasonable request.

\appendix 

\section{Numerical Methods}

In general, it is difficult to find analytical solutions to nonlocal NLSEs, so we use numerical techniques to explore its dynamics and stationary states. The general solution to the NLSE can be written as
\begin{equation}
    \psi(t+dt) = e^{-iHdt}\psi(t),
\end{equation} where $H$ is the Hamiltonian operator. This can be divided into a kinetic energy component $\partial_{x}^2$, which is diagonal in Fourier space, and an effective potential energy component, which is diagonal in real space. Splitting the full evolution operator into $N$ steps corresponding to time intervals $\Delta t=t/N$, the evolution operator for a single time step is
\begin{equation}
    e^{-iH\Delta t}= e^{-iV\Delta t/2}e^{i\partial_{x}^{2}\Delta t}e^{iV\Delta t/2}e^{O(\Delta t^{3})}.
\end{equation}

Numerical time evolution of this type is known as the symmetrized split-step method, which takes advantage of the separable structure of the NLSE by alternating between linear dispersion in Fourier space and nonlinear evolution in real space. In this way, computational expenses are minimized compared to the direct calculation using finite difference methods. Provided that the step size is sufficient small, the error is of order $dt^3$ for a time step of size $dt$ and energy will be conserved to a good approximation \cite{TAHA1984203}.

For beam propagation on a discrete lattice, numerical regularization is essential in order to address numerical instabilities and control spurious high-frequency components, particularly for long time evolutions. We address this by making the substitution $k^2 \rightarrow 2J \cos k$, where $J$ corresponds to the hopping strength of an optical lattice realization. In the long wavelength limit, this reduces to the continuum model $2J \cos k \approx 2J -J k^2$ if $J=1$. Secondly, we smooth the sign nonlinearity term which may be numerically problematic due to the discontinuity by taking $\mathrm{sgn}(\partial_x |\psi|^2) \rightarrow \tanh(\partial_x |\psi|^2/w)$, where $w$ is a cutoff length scale. In the limit of many sites and broad wavepackets compared to $w$, or equivalently the large $w$ limit, this will reduce to the ideal continuum model, which is a purely nonlocal NLSE with a nonlinear term proportional to $\frac{1}{w} \partial_x^2 |\psi|^2$.

We use imaginary time evolution (ITE) to approximate the ground state of our NLSE model. This is a widely used numerical technique which converges a system to its ground state, as long as the system has some initial support on the ground state.  By transforming the real-time evolution $t\rightarrow i\tau$, the time-dependent NLSE becomes a diffusion-like equation that exponentially suppresses higher-energy components of an initial state and thereby projecting it toward the ground state \cite{LEHTOVAARA2007148}.

\bibliography{bibliography.bib}

\begin{thebibliography}{45}%
\makeatletter
\providecommand \@ifxundefined [1]{%
 \@ifx{#1\undefined}
}%
\providecommand \@ifnum [1]{%
 \ifnum #1\expandafter \@firstoftwo
 \else \expandafter \@secondoftwo
 \fi
}%
\providecommand \@ifx [1]{%
 \ifx #1\expandafter \@firstoftwo
 \else \expandafter \@secondoftwo
 \fi
}%
\providecommand \natexlab [1]{#1}%
\providecommand \enquote  [1]{``#1''}%
\providecommand \bibnamefont  [1]{#1}%
\providecommand \bibfnamefont [1]{#1}%
\providecommand \citenamefont [1]{#1}%
\providecommand \href@noop [0]{\@secondoftwo}%
\providecommand \href [0]{\begingroup \@sanitize@url \@href}%
\providecommand \@href[1]{\@@startlink{#1}\@@href}%
\providecommand \@@href[1]{\endgroup#1\@@endlink}%
\providecommand \@sanitize@url [0]{\catcode `\\12\catcode `\$12\catcode `\&12\catcode `\#12\catcode `\^12\catcode `\_12\catcode `\%12\relax}%
\providecommand \@@startlink[1]{}%
\providecommand \@@endlink[0]{}%
\providecommand \url  [0]{\begingroup\@sanitize@url \@url }%
\providecommand \@url [1]{\endgroup\@href {#1}{\urlprefix }}%
\providecommand \urlprefix  [0]{URL }%
\providecommand \Eprint [0]{\href }%
\providecommand \doibase [0]{http://dx.doi.org/}%
\providecommand \selectlanguage [0]{\@gobble}%
\providecommand \bibinfo  [0]{\@secondoftwo}%
\providecommand \bibfield  [0]{\@secondoftwo}%
\providecommand \translation [1]{[#1]}%
\providecommand \BibitemOpen [0]{}%
\providecommand \bibitemStop [0]{}%
\providecommand \bibitemNoStop [0]{.\EOS\space}%
\providecommand \EOS [0]{\spacefactor3000\relax}%
\providecommand \BibitemShut  [1]{\csname bibitem#1\endcsname}%
\let\auto@bib@innerbib\@empty
\bibitem [{\citenamefont {Kr\'olikowski}\ and\ \citenamefont {Bang}(2000)}]{PhysRevE.63.016610}%
  \BibitemOpen
  \bibfield  {author} {\bibinfo {author} {\bibfnamefont {Wies\l{}aw}\ \bibnamefont {Kr\'olikowski}}\ and\ \bibinfo {author} {\bibfnamefont {Ole}\ \bibnamefont {Bang}},\ }\bibfield  {title} {\enquote {\bibinfo {title} {Solitons in nonlocal nonlinear media: Exact solutions},}\ }\href {\doibase 10.1103/PhysRevE.63.016610} {\bibfield  {journal} {\bibinfo  {journal} {Phys. Rev. E}\ }\textbf {\bibinfo {volume} {63}},\ \bibinfo {pages} {016610} (\bibinfo {year} {2000})}\BibitemShut {NoStop}%
\bibitem [{\citenamefont {Krolikowski}\ \emph {et~al.}(2001)\citenamefont {Krolikowski}, \citenamefont {Bang}, \citenamefont {Rasmussen},\ and\ \citenamefont {Wyller}}]{PhysRevE.64.016612}%
  \BibitemOpen
  \bibfield  {author} {\bibinfo {author} {\bibfnamefont {Wieslaw}\ \bibnamefont {Krolikowski}}, \bibinfo {author} {\bibfnamefont {Ole}\ \bibnamefont {Bang}}, \bibinfo {author} {\bibfnamefont {Jens~Juul}\ \bibnamefont {Rasmussen}}, \ and\ \bibinfo {author} {\bibfnamefont {John}\ \bibnamefont {Wyller}},\ }\bibfield  {title} {\enquote {\bibinfo {title} {Modulational instability in nonlocal nonlinear kerr media},}\ }\href {\doibase 10.1103/PhysRevE.64.016612} {\bibfield  {journal} {\bibinfo  {journal} {Phys. Rev. E}\ }\textbf {\bibinfo {volume} {64}},\ \bibinfo {pages} {016612} (\bibinfo {year} {2001})}\BibitemShut {NoStop}%
\bibitem [{\citenamefont {Królikowski}\ \emph {et~al.}(2004)\citenamefont {Królikowski}, \citenamefont {Bang}, \citenamefont {Nikolov}, \citenamefont {Neshev}, \citenamefont {Wyller}, \citenamefont {Rasmussen},\ and\ \citenamefont {Edmundson}}]{Krolikowski2004}%
  \BibitemOpen
  \bibfield  {author} {\bibinfo {author} {\bibfnamefont {W}~\bibnamefont {Królikowski}}, \bibinfo {author} {\bibfnamefont {O}~\bibnamefont {Bang}}, \bibinfo {author} {\bibfnamefont {N~I}\ \bibnamefont {Nikolov}}, \bibinfo {author} {\bibfnamefont {D}~\bibnamefont {Neshev}}, \bibinfo {author} {\bibfnamefont {J}~\bibnamefont {Wyller}}, \bibinfo {author} {\bibfnamefont {J~J}\ \bibnamefont {Rasmussen}}, \ and\ \bibinfo {author} {\bibfnamefont {D}~\bibnamefont {Edmundson}},\ }\bibfield  {title} {\enquote {\bibinfo {title} {Modulational instability, solitons and beam propagation in spatially nonlocal nonlinear media},}\ }\href {\doibase 10.1088/1464-4266/6/5/017} {\bibfield  {journal} {\bibinfo  {journal} {Journal of Optics B: Quantum and Semiclassical Optics}\ }\textbf {\bibinfo {volume} {6}},\ \bibinfo {pages} {S288} (\bibinfo {year} {2004})}\BibitemShut {NoStop}%
\bibitem [{\citenamefont {Turitsyn}\ and\ \citenamefont {Mikhailov}(2002)}]{TURITSYN20021741}%
  \BibitemOpen
  \bibfield  {author} {\bibinfo {author} {\bibfnamefont {S.K.}\ \bibnamefont {Turitsyn}}\ and\ \bibinfo {author} {\bibfnamefont {A.V.}\ \bibnamefont {Mikhailov}},\ }\bibfield  {title} {\enquote {\bibinfo {title} {Chapter 6.2.5 - {A}pplications of {S}olitons},}\ }in\ \href {\doibase https://doi.org/10.1016/B978-012613760-6/50098-X} {\emph {\bibinfo {booktitle} {Scattering}}},\ \bibinfo {editor} {edited by\ \bibinfo {editor} {\bibfnamefont {R.}~\bibnamefont {Pike}}\ and\ \bibinfo {editor} {\bibfnamefont {P.}~\bibnamefont {Sabatier}}}\ (\bibinfo  {publisher} {Academic Press},\ \bibinfo {address} {London},\ \bibinfo {year} {2002})\ pp.\ \bibinfo {pages} {1741--1753}\BibitemShut {NoStop}%
\bibitem [{\citenamefont {Porsezian}\ and\ \citenamefont {Kuriakose}(2003)}]{porsezian2003optical}%
  \BibitemOpen
  \bibfield  {author} {\bibinfo {author} {\bibfnamefont {Kuppuswamy}\ \bibnamefont {Porsezian}}\ and\ \bibinfo {author} {\bibfnamefont {Valakkattil~Chako}\ \bibnamefont {Kuriakose}},\ }\href@noop {} {\emph {\bibinfo {title} {Optical Solitons: Theoretical and Experimental Challenges}}},\ Vol.\ \bibinfo {volume} {613}\ (\bibinfo  {publisher} {Springer Science \& Business Media},\ \bibinfo {year} {2003})\BibitemShut {NoStop}%
\bibitem [{\citenamefont {Ginsberg}\ \emph {et~al.}(2005)\citenamefont {Ginsberg}, \citenamefont {Brand},\ and\ \citenamefont {Hau}}]{PhysRevLett.94.040403}%
  \BibitemOpen
  \bibfield  {author} {\bibinfo {author} {\bibfnamefont {Naomi~S.}\ \bibnamefont {Ginsberg}}, \bibinfo {author} {\bibfnamefont {Joachim}\ \bibnamefont {Brand}}, \ and\ \bibinfo {author} {\bibfnamefont {Lene~Vestergaard}\ \bibnamefont {Hau}},\ }\bibfield  {title} {\enquote {\bibinfo {title} {Observation of hybrid soliton vortex-ring structures in {B}ose-{E}instein condensates},}\ }\href {\doibase 10.1103/PhysRevLett.94.040403} {\bibfield  {journal} {\bibinfo  {journal} {Phys. Rev. Lett.}\ }\textbf {\bibinfo {volume} {94}},\ \bibinfo {pages} {040403} (\bibinfo {year} {2005})}\BibitemShut {NoStop}%
\bibitem [{\citenamefont {Porkolab}\ and\ \citenamefont {Goldman}(1976)}]{porkolab1976upper}%
  \BibitemOpen
  \bibfield  {author} {\bibinfo {author} {\bibfnamefont {M}~\bibnamefont {Porkolab}}\ and\ \bibinfo {author} {\bibfnamefont {MV}~\bibnamefont {Goldman}},\ }\bibfield  {title} {\enquote {\bibinfo {title} {Upper-hybrid solitons and oscillating-two-stream instabilities},}\ }\href@noop {} {\bibfield  {journal} {\bibinfo  {journal} {Phys. Fluids}\ }\textbf {\bibinfo {volume} {19}} (\bibinfo {year} {1976})}\BibitemShut {NoStop}%
\bibitem [{\citenamefont {Montes}(1979)}]{10.1007/3-540-09246-3_9}%
  \BibitemOpen
  \bibfield  {author} {\bibinfo {author} {\bibfnamefont {Carlos}\ \bibnamefont {Montes}},\ }\bibfield  {title} {\enquote {\bibinfo {title} {Nonlinear kinetic equation in plasma physics leading to soliton structures},}\ }in\ \href@noop {} {\emph {\bibinfo {booktitle} {Nonlinear Problems in Theoretical Physics}}},\ \bibinfo {editor} {edited by\ \bibinfo {editor} {\bibfnamefont {A.~F.}\ \bibnamefont {Ra{\~{n}}ada}}}\ (\bibinfo  {publisher} {Springer Berlin Heidelberg},\ \bibinfo {address} {Berlin, Heidelberg},\ \bibinfo {year} {1979})\ pp.\ \bibinfo {pages} {205--216}\BibitemShut {NoStop}%
\bibitem [{\citenamefont {Ankiewicz}\ \emph {et~al.}(2014)\citenamefont {Ankiewicz}, \citenamefont {Wang}, \citenamefont {Wabnitz},\ and\ \citenamefont {Akhmediev}}]{PhysRevE.89.012907}%
  \BibitemOpen
  \bibfield  {author} {\bibinfo {author} {\bibfnamefont {Adrian}\ \bibnamefont {Ankiewicz}}, \bibinfo {author} {\bibfnamefont {Yan}\ \bibnamefont {Wang}}, \bibinfo {author} {\bibfnamefont {Stefan}\ \bibnamefont {Wabnitz}}, \ and\ \bibinfo {author} {\bibfnamefont {Nail}\ \bibnamefont {Akhmediev}},\ }\bibfield  {title} {\enquote {\bibinfo {title} {Extended nonlinear {S}chr\"odinger equation with higher-order odd and even terms and its rogue wave solutions},}\ }\href {\doibase 10.1103/PhysRevE.89.012907} {\bibfield  {journal} {\bibinfo  {journal} {Phys. Rev. E}\ }\textbf {\bibinfo {volume} {89}},\ \bibinfo {pages} {012907} (\bibinfo {year} {2014})}\BibitemShut {NoStop}%
\bibitem [{\citenamefont {Li}\ \emph {et~al.}(2005)\citenamefont {Li}, \citenamefont {Li}, \citenamefont {Malomed}, \citenamefont {Mihalache},\ and\ \citenamefont {Liu}}]{PhysRevA.72.033611}%
  \BibitemOpen
  \bibfield  {author} {\bibinfo {author} {\bibfnamefont {Lu}~\bibnamefont {Li}}, \bibinfo {author} {\bibfnamefont {Zaidong}\ \bibnamefont {Li}}, \bibinfo {author} {\bibfnamefont {Boris~A.}\ \bibnamefont {Malomed}}, \bibinfo {author} {\bibfnamefont {Dumitru}\ \bibnamefont {Mihalache}}, \ and\ \bibinfo {author} {\bibfnamefont {W.~M.}\ \bibnamefont {Liu}},\ }\bibfield  {title} {\enquote {\bibinfo {title} {Exact soliton solutions and nonlinear modulation instability in spinor {B}ose-{E}instein condensates},}\ }\href {\doibase 10.1103/PhysRevA.72.033611} {\bibfield  {journal} {\bibinfo  {journal} {Phys. Rev. A}\ }\textbf {\bibinfo {volume} {72}},\ \bibinfo {pages} {033611} (\bibinfo {year} {2005})}\BibitemShut {NoStop}%
\bibitem [{\citenamefont {Bang}\ \emph {et~al.}(2002)\citenamefont {Bang}, \citenamefont {Krolikowski}, \citenamefont {Wyller},\ and\ \citenamefont {Rasmussen}}]{PhysRevE.66.046619}%
  \BibitemOpen
  \bibfield  {author} {\bibinfo {author} {\bibfnamefont {Ole}\ \bibnamefont {Bang}}, \bibinfo {author} {\bibfnamefont {Wieslaw}\ \bibnamefont {Krolikowski}}, \bibinfo {author} {\bibfnamefont {John}\ \bibnamefont {Wyller}}, \ and\ \bibinfo {author} {\bibfnamefont {Jens~Juul}\ \bibnamefont {Rasmussen}},\ }\bibfield  {title} {\enquote {\bibinfo {title} {Collapse arrest and soliton stabilization in nonlocal nonlinear media},}\ }\href {\doibase 10.1103/PhysRevE.66.046619} {\bibfield  {journal} {\bibinfo  {journal} {Phys. Rev. E}\ }\textbf {\bibinfo {volume} {66}},\ \bibinfo {pages} {046619} (\bibinfo {year} {2002})}\BibitemShut {NoStop}%
\bibitem [{\citenamefont {Ye}\ \emph {et~al.}(2008)\citenamefont {Ye}, \citenamefont {Kartashov},\ and\ \citenamefont {Torner}}]{PhysRevA.77.043821}%
  \BibitemOpen
  \bibfield  {author} {\bibinfo {author} {\bibfnamefont {Fangwei}\ \bibnamefont {Ye}}, \bibinfo {author} {\bibfnamefont {Yaroslav~V.}\ \bibnamefont {Kartashov}}, \ and\ \bibinfo {author} {\bibfnamefont {Lluis}\ \bibnamefont {Torner}},\ }\bibfield  {title} {\enquote {\bibinfo {title} {Stabilization of dipole solitons in nonlocal nonlinear media},}\ }\href {\doibase 10.1103/PhysRevA.77.043821} {\bibfield  {journal} {\bibinfo  {journal} {Phys. Rev. A}\ }\textbf {\bibinfo {volume} {77}},\ \bibinfo {pages} {043821} (\bibinfo {year} {2008})}\BibitemShut {NoStop}%
\bibitem [{\citenamefont {Xu}\ \emph {et~al.}(2006)\citenamefont {Xu}, \citenamefont {Kartashov},\ and\ \citenamefont {Torner}}]{PhysRevE.73.055601}%
  \BibitemOpen
  \bibfield  {author} {\bibinfo {author} {\bibfnamefont {Zhiyong}\ \bibnamefont {Xu}}, \bibinfo {author} {\bibfnamefont {Yaroslav~V.}\ \bibnamefont {Kartashov}}, \ and\ \bibinfo {author} {\bibfnamefont {Lluis}\ \bibnamefont {Torner}},\ }\bibfield  {title} {\enquote {\bibinfo {title} {Stabilization of vector soliton complexes in nonlocal nonlinear media},}\ }\href {\doibase 10.1103/PhysRevE.73.055601} {\bibfield  {journal} {\bibinfo  {journal} {Phys. Rev. E}\ }\textbf {\bibinfo {volume} {73}},\ \bibinfo {pages} {055601} (\bibinfo {year} {2006})}\BibitemShut {NoStop}%
\bibitem [{\citenamefont {Rabie}\ \emph {et~al.}(2024)\citenamefont {Rabie}, \citenamefont {Ahmed}, \citenamefont {Samir},\ and\ \citenamefont {Alnahhass}}]{RABIE2024107589}%
  \BibitemOpen
  \bibfield  {author} {\bibinfo {author} {\bibfnamefont {Wafaa~B.}\ \bibnamefont {Rabie}}, \bibinfo {author} {\bibfnamefont {Hamdy~M.}\ \bibnamefont {Ahmed}}, \bibinfo {author} {\bibfnamefont {Islam}\ \bibnamefont {Samir}}, \ and\ \bibinfo {author} {\bibfnamefont {Mahmoud}\ \bibnamefont {Alnahhass}},\ }\bibfield  {title} {\enquote {\bibinfo {title} {Optical solitons and stability analysis for {NLSE} with nonlocal nonlinearity, nonlinear chromatic dispersion and {K}udryashov’s generalized quintuple-power nonlinearity},}\ }\href {\doibase https://doi.org/10.1016/j.rinp.2024.107589} {\bibfield  {journal} {\bibinfo  {journal} {Results in Physics}\ }\textbf {\bibinfo {volume} {59}},\ \bibinfo {pages} {107589} (\bibinfo {year} {2024})}\BibitemShut {NoStop}%
\bibitem [{\citenamefont {Edmonds}\ \emph {et~al.}(2013)\citenamefont {Edmonds}, \citenamefont {Valiente}, \citenamefont {Juzeliūnas}, \citenamefont {Santos},\ and\ \citenamefont {Ohberg}}]{edmonds_2013}%
  \BibitemOpen
  \bibfield  {author} {\bibinfo {author} {\bibfnamefont {Matthew}\ \bibnamefont {Edmonds}}, \bibinfo {author} {\bibfnamefont {Manuel}\ \bibnamefont {Valiente}}, \bibinfo {author} {\bibfnamefont {Gediminas}\ \bibnamefont {Juzeliūnas}}, \bibinfo {author} {\bibfnamefont {Luis}\ \bibnamefont {Santos}}, \ and\ \bibinfo {author} {\bibfnamefont {Patrik}\ \bibnamefont {Ohberg}},\ }\bibfield  {title} {\enquote {\bibinfo {title} {Simulating an interacting gauge theory with ultracold {B}ose gases},}\ }\href {\doibase 10.1103/PhysRevLett.110.085301} {\bibfield  {journal} {\bibinfo  {journal} {Physical Review Letters}\ }\textbf {\bibinfo {volume} {110}},\ \bibinfo {pages} {085301} (\bibinfo {year} {2013})}\BibitemShut {NoStop}%
\bibitem [{\citenamefont {Zheng}\ \emph {et~al.}(2015)\citenamefont {Zheng}, \citenamefont {Xiong}, \citenamefont {Juzeli\ifmmode~\bar{u}\else \={u}\fi{}nas},\ and\ \citenamefont {Wang}}]{PhysRevA.92.013604}%
  \BibitemOpen
  \bibfield  {author} {\bibinfo {author} {\bibfnamefont {Jun-Hui}\ \bibnamefont {Zheng}}, \bibinfo {author} {\bibfnamefont {Bo}~\bibnamefont {Xiong}}, \bibinfo {author} {\bibfnamefont {Gediminas}\ \bibnamefont {Juzeli\ifmmode~\bar{u}\else \={u}\fi{}nas}}, \ and\ \bibinfo {author} {\bibfnamefont {Daw-Wei}\ \bibnamefont {Wang}},\ }\bibfield  {title} {\enquote {\bibinfo {title} {Topological condensate in an interaction-induced gauge potential},}\ }\href {\doibase 10.1103/PhysRevA.92.013604} {\bibfield  {journal} {\bibinfo  {journal} {Phys. Rev. A}\ }\textbf {\bibinfo {volume} {92}},\ \bibinfo {pages} {013604} (\bibinfo {year} {2015})}\BibitemShut {NoStop}%
\bibitem [{\citenamefont {Biswas}(2009)}]{BISWAS20092845}%
  \BibitemOpen
  \bibfield  {author} {\bibinfo {author} {\bibfnamefont {Anjan}\ \bibnamefont {Biswas}},\ }\bibfield  {title} {\enquote {\bibinfo {title} {Topological 1-soliton solution of the nonlinear {S}chrodinger’s equation with {K}err law nonlinearity in 1+2 dimensions},}\ }\href {\doibase https://doi.org/10.1016/j.cnsns.2008.09.025} {\bibfield  {journal} {\bibinfo  {journal} {Communications in Nonlinear Science and Numerical Simulation}\ }\textbf {\bibinfo {volume} {14}},\ \bibinfo {pages} {2845--2847} (\bibinfo {year} {2009})}\BibitemShut {NoStop}%
\bibitem [{\citenamefont {Faugno}\ \emph {et~al.}(2024)\citenamefont {Faugno}, \citenamefont {Salerno},\ and\ \citenamefont {Ozawa}}]{PhysRevLett.132.023401}%
  \BibitemOpen
  \bibfield  {author} {\bibinfo {author} {\bibfnamefont {W.~N.}\ \bibnamefont {Faugno}}, \bibinfo {author} {\bibfnamefont {Mario}\ \bibnamefont {Salerno}}, \ and\ \bibinfo {author} {\bibfnamefont {Tomoki}\ \bibnamefont {Ozawa}},\ }\bibfield  {title} {\enquote {\bibinfo {title} {Density dependent gauge field inducing emergent {S}u-{S}chrieffer-{H}eeger physics, solitons, and condensates in a discrete nonlinear {S}chr\"odinger equation},}\ }\href {\doibase 10.1103/PhysRevLett.132.023401} {\bibfield  {journal} {\bibinfo  {journal} {Phys. Rev. Lett.}\ }\textbf {\bibinfo {volume} {132}},\ \bibinfo {pages} {023401} (\bibinfo {year} {2024})}\BibitemShut {NoStop}%
\bibitem [{\citenamefont {Bhat}\ \emph {et~al.}(2021)\citenamefont {Bhat}, \citenamefont {Sivaprakasam},\ and\ \citenamefont {Malomed}}]{PhysRevE.103.032206}%
  \BibitemOpen
  \bibfield  {author} {\bibinfo {author} {\bibfnamefont {Ishfaq~Ahmad}\ \bibnamefont {Bhat}}, \bibinfo {author} {\bibfnamefont {S.}~\bibnamefont {Sivaprakasam}}, \ and\ \bibinfo {author} {\bibfnamefont {Boris~A.}\ \bibnamefont {Malomed}},\ }\bibfield  {title} {\enquote {\bibinfo {title} {Modulational instability and soliton generation in chiral {B}ose-{E}instein condensates with zero-energy nonlinearity},}\ }\href {\doibase 10.1103/PhysRevE.103.032206} {\bibfield  {journal} {\bibinfo  {journal} {Phys. Rev. E}\ }\textbf {\bibinfo {volume} {103}},\ \bibinfo {pages} {032206} (\bibinfo {year} {2021})}\BibitemShut {NoStop}%
\bibitem [{\citenamefont {Kai-Xuan~Yao}\ and\ \citenamefont {Chin}(2022)}]{YAO2022}%
  \BibitemOpen
  \bibfield  {author} {\bibinfo {author} {\bibfnamefont {Zhendong~Zhang}\ \bibnamefont {Kai-Xuan~Yao}}\ and\ \bibinfo {author} {\bibfnamefont {Cheng}\ \bibnamefont {Chin}},\ }\bibfield  {title} {\enquote {\bibinfo {title} {Domain-wall dynamics in {B}ose–{E}instein condensates with synthetic gauge fields},}\ }\href {\doibase https://doi.org/10.1038/s41586-021-04250-3} {\bibfield  {journal} {\bibinfo  {journal} {Nature}\ }\textbf {\bibinfo {volume} {602}},\ \bibinfo {pages} {68--72} (\bibinfo {year} {2022})}\BibitemShut {NoStop}%
\bibitem [{\citenamefont {Clark}\ \emph {et~al.}(2018)\citenamefont {Clark}, \citenamefont {Anderson}, \citenamefont {Feng}, \citenamefont {Gaj}, \citenamefont {Levin},\ and\ \citenamefont {Chin}}]{PhysRevLett.121.030402}%
  \BibitemOpen
  \bibfield  {author} {\bibinfo {author} {\bibfnamefont {Logan~W.}\ \bibnamefont {Clark}}, \bibinfo {author} {\bibfnamefont {Brandon~M.}\ \bibnamefont {Anderson}}, \bibinfo {author} {\bibfnamefont {Lei}\ \bibnamefont {Feng}}, \bibinfo {author} {\bibfnamefont {Anita}\ \bibnamefont {Gaj}}, \bibinfo {author} {\bibfnamefont {K.}~\bibnamefont {Levin}}, \ and\ \bibinfo {author} {\bibfnamefont {Cheng}\ \bibnamefont {Chin}},\ }\bibfield  {title} {\enquote {\bibinfo {title} {Observation of density-dependent gauge fields in a {B}ose-{E}instein condensate based on micromotion control in a shaken two-dimensional lattice},}\ }\href {\doibase 10.1103/PhysRevLett.121.030402} {\bibfield  {journal} {\bibinfo  {journal} {Phys. Rev. Lett.}\ }\textbf {\bibinfo {volume} {121}},\ \bibinfo {pages} {030402} (\bibinfo {year} {2018})}\BibitemShut {NoStop}%
\bibitem [{\citenamefont {Martinez}\ \emph {et~al.}(2016)\citenamefont {Martinez}, \citenamefont {Muschik},\ and\ \citenamefont {Schindler}}]{Martinez_2016}%
  \BibitemOpen
  \bibfield  {author} {\bibinfo {author} {\bibfnamefont {E.}~\bibnamefont {Martinez}}, \bibinfo {author} {\bibfnamefont {C.}~\bibnamefont {Muschik}}, \ and\ \bibinfo {author} {\bibfnamefont {P.}~\bibnamefont {Schindler}},\ }\bibfield  {title} {\enquote {\bibinfo {title} {Real-time dynamics of lattice gauge theories with a few-qubit quantum computer},}\ }\href {\doibase https://doi.org/10.1038/nature18318} {\bibfield  {journal} {\bibinfo  {journal} {Nature}\ }\textbf {\bibinfo {volume} {534}},\ \bibinfo {pages} {516--519} (\bibinfo {year} {2016})}\BibitemShut {NoStop}%
\bibitem [{\citenamefont {Görg}\ \emph {et~al.}(2019)\citenamefont {Görg}, \citenamefont {Sandholzer},\ and\ \citenamefont {Minguzzi}}]{Gorg_2019}%
  \BibitemOpen
  \bibfield  {author} {\bibinfo {author} {\bibfnamefont {F.}~\bibnamefont {Görg}}, \bibinfo {author} {\bibfnamefont {K.}~\bibnamefont {Sandholzer}}, \ and\ \bibinfo {author} {\bibfnamefont {J.}~\bibnamefont {Minguzzi}},\ }\bibfield  {title} {\enquote {\bibinfo {title} {Realization of density-dependent {P}eierls phases to engineer quantized gauge fields coupled to ultracold matter},}\ }\href {\doibase https://doi.org/10.1038/s41567-019-0615-4} {\bibfield  {journal} {\bibinfo  {journal} {Nat. Phys.}\ }\textbf {\bibinfo {volume} {15}},\ \bibinfo {pages} {1161--1167} (\bibinfo {year} {2019})}\BibitemShut {NoStop}%
\bibitem [{\citenamefont {Gordon}\ \emph {et~al.}(2007)\citenamefont {Gordon}, \citenamefont {Blakely},\ and\ \citenamefont {Sinton}}]{PhysRevA.75.055801}%
  \BibitemOpen
  \bibfield  {author} {\bibinfo {author} {\bibfnamefont {R.}~\bibnamefont {Gordon}}, \bibinfo {author} {\bibfnamefont {J.~T.}\ \bibnamefont {Blakely}}, \ and\ \bibinfo {author} {\bibfnamefont {D.}~\bibnamefont {Sinton}},\ }\bibfield  {title} {\enquote {\bibinfo {title} {Particle-optical self-trapping},}\ }\href {\doibase 10.1103/PhysRevA.75.055801} {\bibfield  {journal} {\bibinfo  {journal} {Phys. Rev. A}\ }\textbf {\bibinfo {volume} {75}},\ \bibinfo {pages} {055801} (\bibinfo {year} {2007})}\BibitemShut {NoStop}%
\bibitem [{\citenamefont {Lamhot}\ \emph {et~al.}(2010)\citenamefont {Lamhot}, \citenamefont {Barak}, \citenamefont {Peleg},\ and\ \citenamefont {Segev}}]{PhysRevLett.105.163906}%
  \BibitemOpen
  \bibfield  {author} {\bibinfo {author} {\bibfnamefont {Yuval}\ \bibnamefont {Lamhot}}, \bibinfo {author} {\bibfnamefont {Assaf}\ \bibnamefont {Barak}}, \bibinfo {author} {\bibfnamefont {Or}~\bibnamefont {Peleg}}, \ and\ \bibinfo {author} {\bibfnamefont {Mordechai}\ \bibnamefont {Segev}},\ }\bibfield  {title} {\enquote {\bibinfo {title} {Self-trapping of optical beams through thermophoresis},}\ }\href {\doibase 10.1103/PhysRevLett.105.163906} {\bibfield  {journal} {\bibinfo  {journal} {Phys. Rev. Lett.}\ }\textbf {\bibinfo {volume} {105}},\ \bibinfo {pages} {163906} (\bibinfo {year} {2010})}\BibitemShut {NoStop}%
\bibitem [{\citenamefont {Gautam}\ \emph {et~al.}(2019)\citenamefont {Gautam}, \citenamefont {Xiang}, \citenamefont {Lamstein}, \citenamefont {Liang}, \citenamefont {Bezryadina}, \citenamefont {Liang}, \citenamefont {Hansson}, \citenamefont {Wetzel}, \citenamefont {Preece}, \citenamefont {White}, \citenamefont {Silverman}, \citenamefont {Kazarian}, \citenamefont {Xu}, \citenamefont {Morandotti},\ and\ \citenamefont {Chen}}]{Gautam2019}%
  \BibitemOpen
  \bibfield  {author} {\bibinfo {author} {\bibfnamefont {Rekha}\ \bibnamefont {Gautam}}, \bibinfo {author} {\bibfnamefont {Yinxiao}\ \bibnamefont {Xiang}}, \bibinfo {author} {\bibfnamefont {Josh}\ \bibnamefont {Lamstein}}, \bibinfo {author} {\bibfnamefont {Yi}~\bibnamefont {Liang}}, \bibinfo {author} {\bibfnamefont {Anna}\ \bibnamefont {Bezryadina}}, \bibinfo {author} {\bibfnamefont {Guo}\ \bibnamefont {Liang}}, \bibinfo {author} {\bibfnamefont {Tobias}\ \bibnamefont {Hansson}}, \bibinfo {author} {\bibfnamefont {Benjamin}\ \bibnamefont {Wetzel}}, \bibinfo {author} {\bibfnamefont {Daryl}\ \bibnamefont {Preece}}, \bibinfo {author} {\bibfnamefont {Adam}\ \bibnamefont {White}}, \bibinfo {author} {\bibfnamefont {Matthew}\ \bibnamefont {Silverman}}, \bibinfo {author} {\bibfnamefont {Susan}\ \bibnamefont {Kazarian}}, \bibinfo {author} {\bibfnamefont {Jingjun}\ \bibnamefont {Xu}}, \bibinfo {author} {\bibfnamefont {Roberto}\ \bibnamefont {Morandotti}}, \ and\ \bibinfo {author} {\bibfnamefont {Zhigang}\ \bibnamefont
  {Chen}},\ }\bibfield  {title} {\enquote {\bibinfo {title} {Optical force-induced nonlinearity and self-guiding of light in human red blood cell suspensions},}\ }\href {\doibase 10.1038/s41377-019-0142-1} {\bibfield  {journal} {\bibinfo  {journal} {Light: Science {\&} Applications}\ }\textbf {\bibinfo {volume} {8}},\ \bibinfo {pages} {31} (\bibinfo {year} {2019})}\BibitemShut {NoStop}%
\bibitem [{\citenamefont {Chazal}\ \emph {et~al.}(2013)\citenamefont {Chazal}, \citenamefont {Guibas}, \citenamefont {Oudot},\ and\ \citenamefont {Skraba}}]{10.1145/2535927}%
  \BibitemOpen
  \bibfield  {author} {\bibinfo {author} {\bibfnamefont {Fr\'{e}d\'{e}ric}\ \bibnamefont {Chazal}}, \bibinfo {author} {\bibfnamefont {Leonidas~J.}\ \bibnamefont {Guibas}}, \bibinfo {author} {\bibfnamefont {Steve~Y.}\ \bibnamefont {Oudot}}, \ and\ \bibinfo {author} {\bibfnamefont {Primoz}\ \bibnamefont {Skraba}},\ }\bibfield  {title} {\enquote {\bibinfo {title} {Persistence-based clustering in {R}iemannian manifolds},}\ }\href {\doibase 10.1145/2535927} {\bibfield  {journal} {\bibinfo  {journal} {J. ACM}\ }\textbf {\bibinfo {volume} {60}} (\bibinfo {year} {2013}),\ 10.1145/2535927}\BibitemShut {NoStop}%
\bibitem [{\citenamefont {Murugan}\ and\ \citenamefont {Robertson}(2019)}]{murugan2019introductiontopologicaldataanalysis}%
  \BibitemOpen
  \bibfield  {author} {\bibinfo {author} {\bibfnamefont {Jeff}\ \bibnamefont {Murugan}}\ and\ \bibinfo {author} {\bibfnamefont {Duncan}\ \bibnamefont {Robertson}},\ }\href {https://arxiv.org/abs/1904.11044} {\enquote {\bibinfo {title} {An introduction to topological data analysis for physicists: From {LGM} to {FRB}s},}\ } (\bibinfo {year} {2019}),\ \Eprint {http://arxiv.org/abs/1904.11044} {arXiv:1904.11044 [astro-ph.IM]} \BibitemShut {NoStop}%
\bibitem [{\citenamefont {Leykam}\ and\ \citenamefont {Angelakis}(2023)}]{Leykam31122023}%
  \BibitemOpen
  \bibfield  {author} {\bibinfo {author} {\bibfnamefont {Daniel}\ \bibnamefont {Leykam}}\ and\ \bibinfo {author} {\bibfnamefont {Dimitris~G.}\ \bibnamefont {Angelakis}},\ }\bibfield  {title} {\enquote {\bibinfo {title} {Topological data analysis and machine learning},}\ }\href {\doibase 10.1080/23746149.2023.2202331} {\bibfield  {journal} {\bibinfo  {journal} {Advances in Physics: X}\ }\textbf {\bibinfo {volume} {8}},\ \bibinfo {pages} {2202331} (\bibinfo {year} {2023})},\ \Eprint {http://arxiv.org/abs/https://doi.org/10.1080/23746149.2023.2202331} {https://doi.org/10.1080/23746149.2023.2202331} \BibitemShut {NoStop}%
\bibitem [{\citenamefont {Olsthoorn}(2023)}]{PhysRevB.107.115174}%
  \BibitemOpen
  \bibfield  {author} {\bibinfo {author} {\bibfnamefont {Bart}\ \bibnamefont {Olsthoorn}},\ }\bibfield  {title} {\enquote {\bibinfo {title} {Persistent homology of quantum entanglement},}\ }\href {\doibase 10.1103/PhysRevB.107.115174} {\bibfield  {journal} {\bibinfo  {journal} {Phys. Rev. B}\ }\textbf {\bibinfo {volume} {107}},\ \bibinfo {pages} {115174} (\bibinfo {year} {2023})}\BibitemShut {NoStop}%
\bibitem [{\citenamefont {Cao}\ \emph {et~al.}(2023)\citenamefont {Cao}, \citenamefont {Leykam},\ and\ \citenamefont {Angelakis}}]{PhysRevE.107.044204}%
  \BibitemOpen
  \bibfield  {author} {\bibinfo {author} {\bibfnamefont {Harvey}\ \bibnamefont {Cao}}, \bibinfo {author} {\bibfnamefont {Daniel}\ \bibnamefont {Leykam}}, \ and\ \bibinfo {author} {\bibfnamefont {Dimitris~G.}\ \bibnamefont {Angelakis}},\ }\bibfield  {title} {\enquote {\bibinfo {title} {Unravelling quantum chaos using persistent homology},}\ }\href {\doibase 10.1103/PhysRevE.107.044204} {\bibfield  {journal} {\bibinfo  {journal} {Phys. Rev. E}\ }\textbf {\bibinfo {volume} {107}},\ \bibinfo {pages} {044204} (\bibinfo {year} {2023})}\BibitemShut {NoStop}%
\bibitem [{\citenamefont {Hamilton}\ and\ \citenamefont {Leditzky}(2024)}]{Hamilton2024}%
  \BibitemOpen
  \bibfield  {author} {\bibinfo {author} {\bibfnamefont {Gregory}\ \bibnamefont {Hamilton}}\ and\ \bibinfo {author} {\bibfnamefont {Felix}\ \bibnamefont {Leditzky}},\ }\bibfield  {title} {\enquote {\bibinfo {title} {Probing multipartite entanglement through persistent homology},}\ }\href {\doibase https://doi.org/10.1007/s00220-024-04953-4} {\bibfield  {journal} {\bibinfo  {journal} {Commun. Math. Phys.}\ }\textbf {\bibinfo {volume} {405}} (\bibinfo {year} {2024}),\ https://doi.org/10.1007/s00220-024-04953-4}\BibitemShut {NoStop}%
\bibitem [{\citenamefont {Donato}\ \emph {et~al.}(2016)\citenamefont {Donato}, \citenamefont {Gori}, \citenamefont {Pettini}, \citenamefont {Petri}, \citenamefont {De~Nigris}, \citenamefont {Franzosi},\ and\ \citenamefont {Vaccarino}}]{PhysRevE.93.052138}%
  \BibitemOpen
  \bibfield  {author} {\bibinfo {author} {\bibfnamefont {Irene}\ \bibnamefont {Donato}}, \bibinfo {author} {\bibfnamefont {Matteo}\ \bibnamefont {Gori}}, \bibinfo {author} {\bibfnamefont {Marco}\ \bibnamefont {Pettini}}, \bibinfo {author} {\bibfnamefont {Giovanni}\ \bibnamefont {Petri}}, \bibinfo {author} {\bibfnamefont {Sarah}\ \bibnamefont {De~Nigris}}, \bibinfo {author} {\bibfnamefont {Roberto}\ \bibnamefont {Franzosi}}, \ and\ \bibinfo {author} {\bibfnamefont {Francesco}\ \bibnamefont {Vaccarino}},\ }\bibfield  {title} {\enquote {\bibinfo {title} {Persistent homology analysis of phase transitions},}\ }\href {\doibase 10.1103/PhysRevE.93.052138} {\bibfield  {journal} {\bibinfo  {journal} {Phys. Rev. E}\ }\textbf {\bibinfo {volume} {93}},\ \bibinfo {pages} {052138} (\bibinfo {year} {2016})}\BibitemShut {NoStop}%
\bibitem [{\citenamefont {D.~Cohen-Steiner}(2010)}]{CohenSteiner2010}%
  \BibitemOpen
  \bibfield  {author} {\bibinfo {author} {\bibfnamefont {J.~Harer}\ \bibnamefont {D.~Cohen-Steiner}, \bibfnamefont {H.~Edelsbrunner}},\ }\bibfield  {title} {\enquote {\bibinfo {title} {Lipschitz functions have l p -stable persistence},}\ }\href {\doibase 10.1007/s10208-010-9060-6} {\bibfield  {journal} {\bibinfo  {journal} {Found. Comput. Math}\ }\textbf {\bibinfo {volume} {10}},\ \bibinfo {pages} {127--139} (\bibinfo {year} {2010})}\BibitemShut {NoStop}%
\bibitem [{\citenamefont {Hamilton}\ \emph {et~al.}(2022)\citenamefont {Hamilton}, \citenamefont {Dore},\ and\ \citenamefont {Plumberg}}]{PhysRevC.106.064912}%
  \BibitemOpen
  \bibfield  {author} {\bibinfo {author} {\bibfnamefont {Greg}\ \bibnamefont {Hamilton}}, \bibinfo {author} {\bibfnamefont {Travis}\ \bibnamefont {Dore}}, \ and\ \bibinfo {author} {\bibfnamefont {Christopher}\ \bibnamefont {Plumberg}},\ }\bibfield  {title} {\enquote {\bibinfo {title} {Applications of persistent homology in nuclear collisions},}\ }\href {\doibase 10.1103/PhysRevC.106.064912} {\bibfield  {journal} {\bibinfo  {journal} {Phys. Rev. C}\ }\textbf {\bibinfo {volume} {106}},\ \bibinfo {pages} {064912} (\bibinfo {year} {2022})}\BibitemShut {NoStop}%
\bibitem [{\citenamefont {Cole}\ \emph {et~al.}(2021)\citenamefont {Cole}, \citenamefont {Loges},\ and\ \citenamefont {Shiu}}]{PhysRevB.104.104426}%
  \BibitemOpen
  \bibfield  {author} {\bibinfo {author} {\bibfnamefont {Alex}\ \bibnamefont {Cole}}, \bibinfo {author} {\bibfnamefont {Gregory~J.}\ \bibnamefont {Loges}}, \ and\ \bibinfo {author} {\bibfnamefont {Gary}\ \bibnamefont {Shiu}},\ }\bibfield  {title} {\enquote {\bibinfo {title} {Quantitative and interpretable order parameters for phase transitions from persistent homology},}\ }\href {\doibase 10.1103/PhysRevB.104.104426} {\bibfield  {journal} {\bibinfo  {journal} {Phys. Rev. B}\ }\textbf {\bibinfo {volume} {104}},\ \bibinfo {pages} {104426} (\bibinfo {year} {2021})}\BibitemShut {NoStop}%
\bibitem [{\citenamefont {Carriere}\ \emph {et~al.}(2021)\citenamefont {Carriere}, \citenamefont {Chazal}, \citenamefont {Glisse}, \citenamefont {Ike}, \citenamefont {Kannan},\ and\ \citenamefont {Umeda}}]{pmlr-v139-carriere21a}%
  \BibitemOpen
  \bibfield  {author} {\bibinfo {author} {\bibfnamefont {Mathieu}\ \bibnamefont {Carriere}}, \bibinfo {author} {\bibfnamefont {Frederic}\ \bibnamefont {Chazal}}, \bibinfo {author} {\bibfnamefont {Marc}\ \bibnamefont {Glisse}}, \bibinfo {author} {\bibfnamefont {Yuichi}\ \bibnamefont {Ike}}, \bibinfo {author} {\bibfnamefont {Hariprasad}\ \bibnamefont {Kannan}}, \ and\ \bibinfo {author} {\bibfnamefont {Yuhei}\ \bibnamefont {Umeda}},\ }\bibfield  {title} {\enquote {\bibinfo {title} {Optimizing persistent homology based functions},}\ }in\ \href {https://proceedings.mlr.press/v139/carriere21a.html} {\emph {\bibinfo {booktitle} {Proceedings of the 38th International Conference on Machine Learning}}},\ \bibinfo {series} {Proc. Mach. Learn. Res.}, Vol.\ \bibinfo {volume} {139}\ (\bibinfo {year} {2021})\ pp.\ \bibinfo {pages} {1294--1303}\BibitemShut {NoStop}%
\bibitem [{\citenamefont {Ma}\ and\ \citenamefont {Huang}(2013)}]{MA2013718}%
  \BibitemOpen
  \bibfield  {author} {\bibinfo {author} {\bibfnamefont {Manjun}\ \bibnamefont {Ma}}\ and\ \bibinfo {author} {\bibfnamefont {Zhe}\ \bibnamefont {Huang}},\ }\bibfield  {title} {\enquote {\bibinfo {title} {Bright soliton solution of a {G}ross–{P}itaevskii equation},}\ }\href {\doibase https://doi.org/10.1016/j.aml.2013.02.002} {\bibfield  {journal} {\bibinfo  {journal} {Applied Mathematics Letters}\ }\textbf {\bibinfo {volume} {26}},\ \bibinfo {pages} {718--724} (\bibinfo {year} {2013})}\BibitemShut {NoStop}%
\bibitem [{\citenamefont {Karjanto}(2022)}]{KARJANTO2022}%
  \BibitemOpen
  \bibfield  {author} {\bibinfo {author} {\bibfnamefont {Natanael}\ \bibnamefont {Karjanto}},\ }\bibfield  {title} {\enquote {\bibinfo {title} {Bright soliton solution of the nonlinear {S}chrödinger equation: Fourier spectrum and fundamental characteristics},}\ }\href {\doibase https://doi.org/10.3390/math10234559} {\bibfield  {journal} {\bibinfo  {journal} {Mathematics}\ }\textbf {\bibinfo {volume} {10}} (\bibinfo {year} {2022}),\ https://doi.org/10.3390/math10234559}\BibitemShut {NoStop}%
\bibitem [{\citenamefont {Lin}\ \emph {et~al.}(2008)\citenamefont {Lin}, \citenamefont {Zhang},\ and\ \citenamefont {Duan}}]{PhysRevA.77.043626}%
  \BibitemOpen
  \bibfield  {author} {\bibinfo {author} {\bibfnamefont {G.-D.}\ \bibnamefont {Lin}}, \bibinfo {author} {\bibfnamefont {Wei}\ \bibnamefont {Zhang}}, \ and\ \bibinfo {author} {\bibfnamefont {L.-M.}\ \bibnamefont {Duan}},\ }\bibfield  {title} {\enquote {\bibinfo {title} {Characteristics of {B}ose-{E}instein condensation in an optical lattice},}\ }\href {\doibase 10.1103/PhysRevA.77.043626} {\bibfield  {journal} {\bibinfo  {journal} {Phys. Rev. A}\ }\textbf {\bibinfo {volume} {77}},\ \bibinfo {pages} {043626} (\bibinfo {year} {2008})}\BibitemShut {NoStop}%
\bibitem [{\citenamefont {Tajiri}\ and\ \citenamefont {Watanabe}(1998)}]{PhysRevE.57.3510}%
  \BibitemOpen
  \bibfield  {author} {\bibinfo {author} {\bibfnamefont {Masayoshi}\ \bibnamefont {Tajiri}}\ and\ \bibinfo {author} {\bibfnamefont {Yosuke}\ \bibnamefont {Watanabe}},\ }\bibfield  {title} {\enquote {\bibinfo {title} {Breather solutions to the focusing nonlinear {S}chr\"odinger equation},}\ }\href {\doibase 10.1103/PhysRevE.57.3510} {\bibfield  {journal} {\bibinfo  {journal} {Phys. Rev. E}\ }\textbf {\bibinfo {volume} {57}},\ \bibinfo {pages} {3510--3519} (\bibinfo {year} {1998})}\BibitemShut {NoStop}%
\bibitem [{\citenamefont {Chen}\ \emph {et~al.}(2025)\citenamefont {Chen}, \citenamefont {Mao}, \citenamefont {Chen}, \citenamefont {Li}, \citenamefont {Tan}, \citenamefont {Li},\ and\ \citenamefont {Zhang}}]{CHEN2025112776}%
  \BibitemOpen
  \bibfield  {author} {\bibinfo {author} {\bibfnamefont {Wenqiang}\ \bibnamefont {Chen}}, \bibinfo {author} {\bibfnamefont {Jianyong}\ \bibnamefont {Mao}}, \bibinfo {author} {\bibfnamefont {Kai}\ \bibnamefont {Chen}}, \bibinfo {author} {\bibfnamefont {Xun}\ \bibnamefont {Li}}, \bibinfo {author} {\bibfnamefont {Yu}~\bibnamefont {Tan}}, \bibinfo {author} {\bibfnamefont {Ming}\ \bibnamefont {Li}}, \ and\ \bibinfo {author} {\bibfnamefont {Lei}\ \bibnamefont {Zhang}},\ }\bibfield  {title} {\enquote {\bibinfo {title} {Generation of high uniformity flat-top beams by reconstructing the amplitude distribution at descending edges},}\ }\href {\doibase https://doi.org/10.1016/j.optlastec.2025.112776} {\bibfield  {journal} {\bibinfo  {journal} {Optics \& Laser Technology}\ }\textbf {\bibinfo {volume} {186}},\ \bibinfo {pages} {112776} (\bibinfo {year} {2025})}\BibitemShut {NoStop}%
\bibitem [{\citenamefont {Zheng}\ \emph {et~al.}(2021)\citenamefont {Zheng}, \citenamefont {Tan}, \citenamefont {Han}, \citenamefont {Ren}, \citenamefont {Wang}, \citenamefont {Zhao},\ and\ \citenamefont {Si}}]{ZHENG2021105016}%
  \BibitemOpen
  \bibfield  {author} {\bibinfo {author} {\bibfnamefont {Yipeng}\ \bibnamefont {Zheng}}, \bibinfo {author} {\bibfnamefont {Wenjiang}\ \bibnamefont {Tan}}, \bibinfo {author} {\bibfnamefont {Dongdong}\ \bibnamefont {Han}}, \bibinfo {author} {\bibfnamefont {Kaili}\ \bibnamefont {Ren}}, \bibinfo {author} {\bibfnamefont {Yongwang}\ \bibnamefont {Wang}}, \bibinfo {author} {\bibfnamefont {Feng}\ \bibnamefont {Zhao}}, \ and\ \bibinfo {author} {\bibfnamefont {Jinhai}\ \bibnamefont {Si}},\ }\bibfield  {title} {\enquote {\bibinfo {title} {Propagation of high-power optical flat-topped beams in strongly nonlinear media},}\ }\href {\doibase https://doi.org/10.1016/j.rinp.2021.105016} {\bibfield  {journal} {\bibinfo  {journal} {Results in Physics}\ }\textbf {\bibinfo {volume} {31}},\ \bibinfo {pages} {105016} (\bibinfo {year} {2021})}\BibitemShut {NoStop}%
\bibitem [{\citenamefont {Taha}\ and\ \citenamefont {Ablowitz}(1984)}]{TAHA1984203}%
  \BibitemOpen
  \bibfield  {author} {\bibinfo {author} {\bibfnamefont {Thiab~R}\ \bibnamefont {Taha}}\ and\ \bibinfo {author} {\bibfnamefont {Mark~I}\ \bibnamefont {Ablowitz}},\ }\bibfield  {title} {\enquote {\bibinfo {title} {Analytical and numerical aspects of certain nonlinear evolution equations. {II}. {N}umerical, nonlinear {S}chrödinger equation},}\ }\href {\doibase https://doi.org/10.1016/0021-9991(84)90003-2} {\bibfield  {journal} {\bibinfo  {journal} {Journal of Computational Physics}\ }\textbf {\bibinfo {volume} {55}},\ \bibinfo {pages} {203--230} (\bibinfo {year} {1984})}\BibitemShut {NoStop}%
\bibitem [{\citenamefont {Lehtovaara}\ \emph {et~al.}(2007)\citenamefont {Lehtovaara}, \citenamefont {Toivanen},\ and\ \citenamefont {Eloranta}}]{LEHTOVAARA2007148}%
  \BibitemOpen
  \bibfield  {author} {\bibinfo {author} {\bibfnamefont {L.}~\bibnamefont {Lehtovaara}}, \bibinfo {author} {\bibfnamefont {J.}~\bibnamefont {Toivanen}}, \ and\ \bibinfo {author} {\bibfnamefont {J.}~\bibnamefont {Eloranta}},\ }\bibfield  {title} {\enquote {\bibinfo {title} {Solution of time-independent {S}chrödinger equation by the imaginary time propagation method},}\ }\href {\doibase https://doi.org/10.1016/j.jcp.2006.06.006} {\bibfield  {journal} {\bibinfo  {journal} {Journal of Computational Physics}\ }\textbf {\bibinfo {volume} {221}},\ \bibinfo {pages} {148--157} (\bibinfo {year} {2007})}\BibitemShut {NoStop}%
\end{thebibliography}%

\end{document}